\documentclass{kluwer}
\usepackage{epsfig}
\usepackage{wrapfig}
\begin{document}

\begin{article}
\begin{opening}         
  \title{Non-ideal MHD Properties of Magnetic Flux Tubes
    in the Solar Photosphere} 
  \author{Jens \surname{Kleimann}}
  \author{Gunnar \surname{Hornig}}  
  \institute{
    Institut f\"ur Theoretische Physik IV,
    Ruhr-Universit\"at Bochum,\\ 44780 Bochum, Germany
  }
\begin{abstract}
  Magnetic flux tubes reaching from the solar convective zone into the
  chromosphere have to pass through the relatively cool, and therefore
  non-ideal (i.e.~resistive) photospheric region enclosed between the highly
  ideal sub-photospheric and chromospheric plasma. It is shown that stationary
  MHD equilibria of magnetic flux tubes which pass through this region require
  an inflow of photospheric material into the flux tube and a deviation from
  iso-rotation along the tube axis. This means that there is a difference in
  angular velocity of the plasma flow inside the tube below and above the
  non-ideal region. Both effects increase with decreasing cross section of the
  tube. Although for characteristic parameters of thick flux tubes the effect
  is negligible, a scaling law indicates its importance for small-scale
  structures. The relevance of this ``inflow effect'' for the expansion of
  flux tubes above the photosphere is discussed.
\end{abstract}
\abbreviations{
  \abbrev{MHD}{magnetohydrodynamics},
  \abbrev{ODE}{ordinary differential equation},
  \abbrev{rhs}{right hand side}
}
\end{opening}

\section{Introduction}
The interaction of solar flux tubes with the surrounding plasma is usually
treated in the framework of ideal MHD (i.e.~with zero resistivity), in which
no exchange of plasma between the flux tube and its environment is possible.
While this approach appears to be well suited for both the convection zone and
the upper chromosphere (where the degree of ionisation is sufficiently high),
it becomes doubtful for the relatively cold and therefore almost neutral
photospheric plasma (see Figure~4). It this resistive layer, deviations from
the rigid coupling between fluid and field must be anticipated. This could
have important consequences for the widely used conception of flux tubes being
wound up by photospheric motions. Also the strict separation of plasma within
the flux tube from its environment as required in ideal MHD might break down
in this resistive layer. The purpose of this work is to compute this deviation
from ideal MHD in a self-consistent manner.\\
First we consider a stationary magnetic flux tube with both ends anchored in
the convective zone (see Figure 2a). The flux tube can be thought of as
consisting of a set of nested tubes which are flux surfaces for the magnetic
field. Assigning to each of these surfaces the magnetic flux it encloses
defines a function  $\Psi({\bf r})$, ${\bf B}({\bf r})\cdot\nabla \Psi=0$,
which is zero on the tube axis and monotonously increases outwards. (We assume
that there are  no field reversals within the flux tube.) In a stationary
situation, any ideal MHD flow has to preserve these flux surfaces, and the
plasma velocity has to be tangential to the surfaces of constant $\Psi$,
${\bf v} \cdot \nabla \Psi=0$. The lower boundary of the domain under
consideration (given by the lower boundary of the photosphere) is a surface
which intersects the flux tube twice. Any plasma motion imposed on the
boundary at one footpoint implies a corresponding motion at the other
footpoint. The exact relation between these motions is derived from ``ideal''
Ohm's law (i.e.~as it is known from ideal MHD)
\begin{equation}
  -\nabla \Phi + {\bf v} \times {\bf B} = 0 \ .
  \label{idOhm}
\end{equation}  
At the lower boundary, the plasma velocity and the magnetic field may be
decomposed into their poloidal and toroidal components:
\begin{equation}
  \label{split_pt}
  {\bf v} = {\bf v}_{\rm t} + {\bf v}_{\rm p} \quad \mbox{and} \quad  {\bf B}
  = {\bf B}_{\rm t} + {\bf  B}_{\rm p} \ .
\end{equation}
\begin{figure}
  \centerline{\includegraphics[width=10cm]{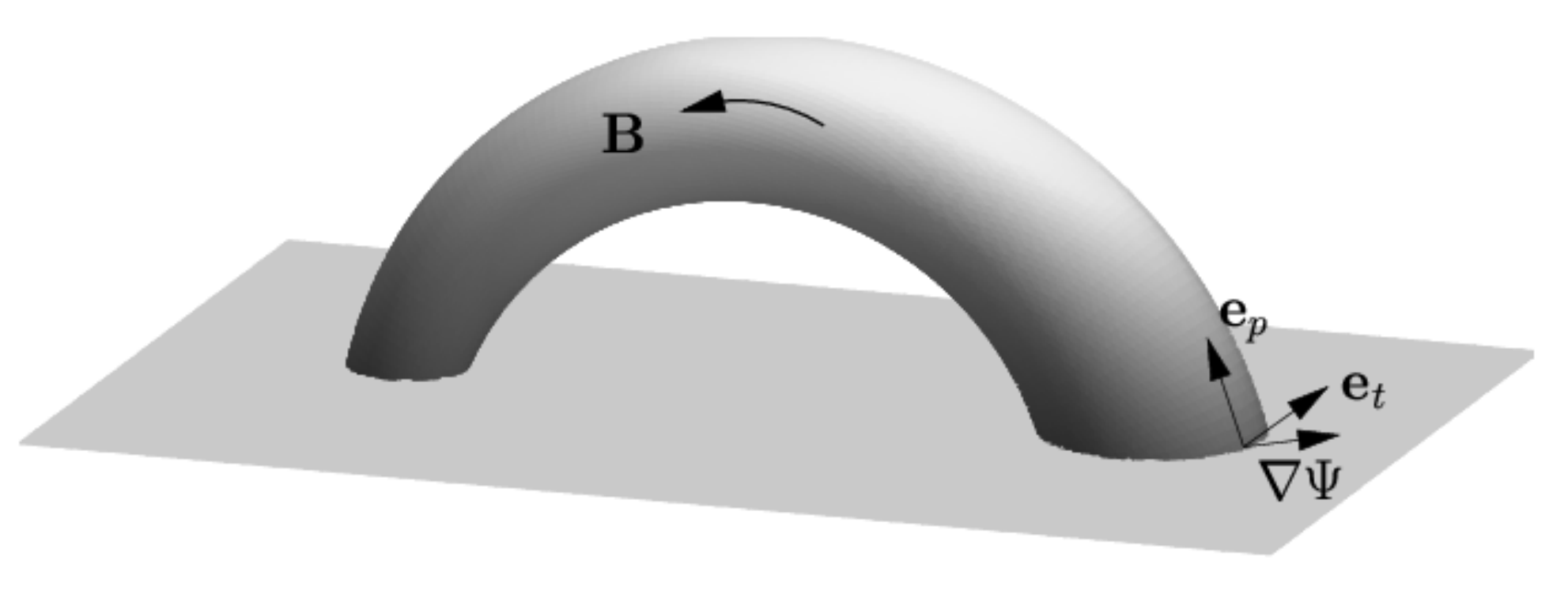}
    \caption{Orientation of unit vectors ${\bf e}_{\rm p}$ and ${\bf e}_{\rm t}$
      with respect to the tube surface and the lower photospheric boundary.}
  }
\end{figure}
The toroidal components are directed along the intersection of the boundary
surface with the $\Psi$-surfaces of the flux tube. Their orientation can be
defined by requiring that the toroidal unit vector ${\bf e}_{\rm t}$ have a
positive orientation with respect to the magnetic field vector on the tube
axis. The poloidal components are also tangential to the $\Psi$ surfaces but
perpendicular to ${\bf e}_{\rm t}$, with a unit vector
${\bf e}_{\rm p} := \nabla \Psi \times {\bf e}_{\rm t} \ / \ \|\nabla \Psi\|$
(see Figure~1). Assuming ${\bf v}_{\rm p}={\bf 0}$ and using
\mbox{${\bf B}_{\rm p}= b_{\rm p} \nabla \Psi \times {\bf e}_{\rm t}$} and
${\bf v}_{\rm t}= v_{\rm t} \ {\bf e}_{\rm t}$, Equation (\ref{idOhm}) yields
\begin{eqnarray}
  \label{gradeq}
  \nabla \Phi &=& {\bf v}_{\rm t} \times {\bf B}_{\rm p}
  = v_{\rm t} \ b_{\rm p} \ \nabla \Psi\\
  \Rightarrow \quad v_{\rm t} &=& \frac{1}{b_{\rm p}}
  \frac{\partial \Phi}{\partial \Psi}
  \label{polvel}
\end{eqnarray}
since (\ref{gradeq}) implies $\Phi=\Phi(\Psi)$. This equation shows that a
given $v_{\rm t}$ distribution at one end of the flux tube determines
${\partial \Phi}/{\partial \Psi}$, a function which depends only on $\Psi$ and
is thus constant along the flux surfaces, thereby inducing a corresponding
$v_{\rm t}$ distribution at the other end. For a flux tube which is
perpendicular to the boundary and which has circular flux surfaces
$\Psi=\Psi(r)$ (where $r$ is the distance from the tube axis), the poloidal
component of ${\bf B}$ is given by
${\bf B}_{\rm p} = 1/(2\pi r) \ \nabla \Psi(r) \times {\bf e}_{\rm t}$ and hence
$b_{\rm p}= 1/ (2\pi r)$. In this case $v_{\rm t}$ and the angular velocity
$\omega := v_{\rm t}/r$ are functions of $r$ only, i.e.~they are constant on
each flux surface $\Psi$. This is simply Ferraro's law of iso-rotation
\cite{Moff}. In the general case $v_{\rm t}$ is not constant on flux surfaces,
but an integration along ${\bf e}_{\rm t}$ yields the circulation time  
\begin{equation}
  T(\Psi) = \oint{\frac{{\rm d}l}{v_{\rm t}}}
  = \left(\frac{\partial \Phi}{\partial \Psi}\right)^{-1}
  \oint b_{\rm p} \ {\rm d}l
\end{equation}
which only depends on the flux surface. This quantity (or the corresponding
angular velocity $\Omega= 2\pi/T$) explicitly shows the coupling of the
toroidal velocity field between both ends of the flux tube.\\
While the preceding results were based on the idealness of the plasma, we will
now investigate the effect of a non-ideal region the flux tube has to pass.
This non-ideal region is given by the comparatively cold photosphere. Here a
possible slippage effect due to the non-ideal photospheric region would result
in a deviation of $v_{\rm p}$ from (\ref{polvel}). Also in the case of
incompatible poloidal velocities on both footpoints the onset of slippage will
keep the resulting twist of the flux tube finite, as opposed to the infinite
``winding-up'' of field lines expected for ideal MHD.

\section{The model}
To study the effect of a resistive layer on the flux tube, it is sufficient to
consider only one half of the tube and concentrate on the photospheric region
close to the footpoint, as shown in Figure 2a. For simplicity, we will
restrict ourselves to {\em stationary}, axisymmetric solutions. The ensuing
calculations will use cylindrical coordinates $[r,\phi,z]$, with unit vectors
$[{\bf e}_r,{\bf e}_{\phi},{\bf e}_z]$. The $(z=0)$--plane is given by the
photosphere's lower boundary, while the $z$--axis coincides with the tube axis
and is pointing away from the Sun. The problem's axial symmetry is now
conveniently incorporated by setting $\partial_{\phi} = 0$. With
$\partial_t = 0$, the set of MHD equations to be solved for the mass flow
velocity {\bf v} and the fields {\bf B} and ${\bf E} = -\nabla \Phi$ consists
of the momentum balance (\ref{impbil_T}), a resistive Ohm's law (\ref{ohm}),
the equation of continuity (\ref{cont}), and the remaining Maxwell equations
(\ref{maxrotB}, \ref{maxdivB}):   
\begin{eqnarray}
  \label{impbil_T}
  {\bf 0} &=& - \nabla P + {\bf j \times B} + \rho \ {\bf g}\\
  \eta \ \bf j &=& - \nabla \Phi + \bf v \times {\bf B} \label{ohm}\\
  0 &=& \nabla \cdot (\rho \ {\bf v}) \label{cont} \\
  \mu \ {\bf j} &=& \nabla \times {\bf B}  \label{maxrotB} \\
  0 &=& \bf \nabla \cdot {\bf B} \label{maxdivB}
\end{eqnarray}
As usual, $\rho$ and $\eta$ denote the plasma's mass density and resistivity,
respectively. The inertia term  $\rho ({\bf v} \cdot \nabla ) {\bf v}$ is
omitted from (\ref{impbil_T}) since its ratio to the induction term
${\bf j \times B}$ is of order ${\cal O}[(v/v_{\rm A})^2]$, where
\begin{equation}
  v_{\rm A} := B/ \sqrt{\mu \ \rho}
\end{equation}
is the {\em Alfv\'en} velocity.
Adopting $B \approx$ 0.1 T and $\rho \approx 10^{-6}$ kg m$^{-3}$ as
characteristic values for our photospheric flux tube yields $v_{\rm A} \approx$
90 km s$^{-1}$, which is large compared to the magnitude of observed
photospheric plasma motions of $v_{\rm obs} \approx$ 5 km s$^{-1}$.
Section~\ref{subA-flows} gives an {\sl a posteriori} verification of this
conjecture. 

\begin{figure}[h]
  \tabcapfont
  \centerline{
    \begin{tabular}{c@{\hspace{1pc}}c}
      \includegraphics[width=6cm]{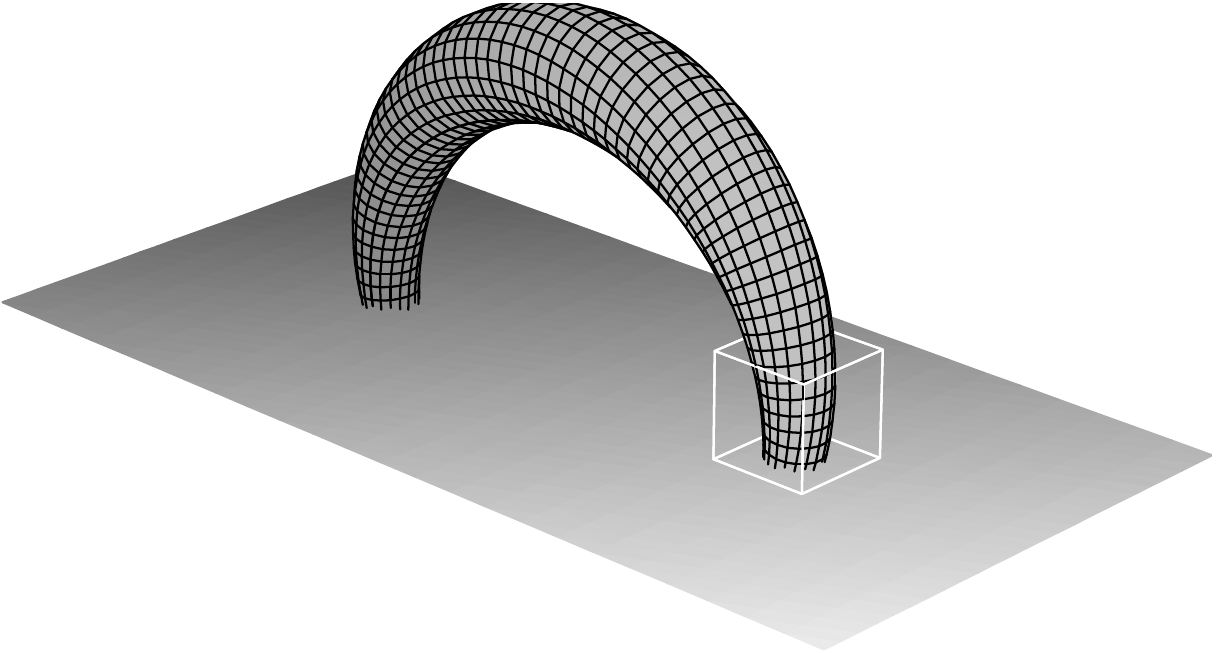} &
      \includegraphics[width=5cm]{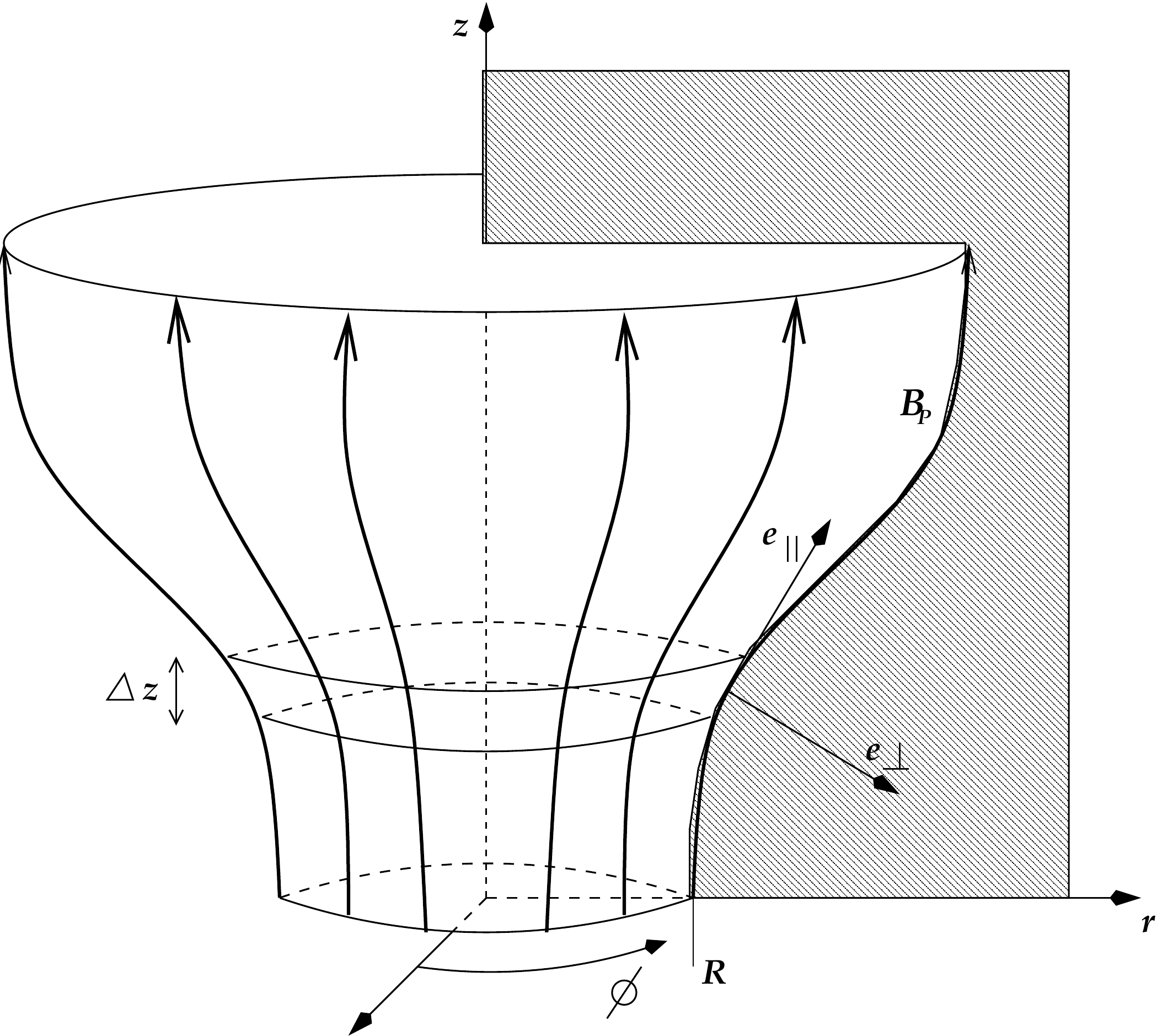} \\
      Figure 2 a. & Figure 2 b. 
    \end{tabular}
    \caption{Left: Flux tube emerging from the Sun's photosphere. The white box
      marks the section on which our computations focus. Right: Our flux tube
      model. The shaded area indicates the poloidal plane in which
      ${\bf B}_{\rm p}$, ${\bf v}_{\perp}$ and ${\bf e}_{\parallel,\perp}$ are
      located.}
  }
\end{figure}

\section{Resistive Inflow}
\label{sect_inflow}
\subsection{Derivation of the Inflow Equation}
An important difference between the ideal and the non-ideal case is the
exchange of plasma between the flux tube and its environment, a process which
is impossible in ideal MHD. The plasma flow across the flux surfaces of the
magnetic field can be derived from Ohm's law (\ref{ohm}) alone
\begin{equation}
  \label{ohm_2}
  -\nabla \Phi+{\bf v \times B} =  {\hat \eta} \ \nabla \times {\bf B}  
\end{equation}
with (\ref{maxrotB}) inserted and ${\hat \eta} :=\eta/ \mu$ substituted. Again
we decompose {\bf v} and {\bf B} similar to (\ref{split_pt}) into their
toroidal and poloidal components, where the toroidal component is directed
along ${\bf e}_{\phi}$ and the poloidal plane is the $r-z$-plane. This yields,
after insertion into (\ref{ohm_2}),
\begin{equation}
  \label{polOhm}
  {\bf v}_{\rm p} \times {\bf B}_{\rm p}
  = {\hat \eta} \ \nabla \times {\bf B}_{\rm p}
\end{equation}
as the poloidal component of Equation (\ref{ohm_2}). Now let
\begin{equation}
  {\bf e}_\parallel := \frac{1}{B_{\rm p}} \ {\bf B}_{\rm p}\\
  \qquad \mbox{and} \qquad
  {\bf e}_{\perp} := {\bf e}_{\rm t} \times {\bf e}_{\parallel}
\end{equation}
be two orthonormal vector fields parallel and perpendicular to
${\bf B}_{\rm p}$. Then the crossproduct of Equation (\ref{ohm_2}) with
${\bf B}_{\rm p}$, together with
${\bf v}_{\rm p} = v_\perp {\bf e}_{\perp} + v_\parallel {\bf e}_\parallel$
yields:
\begin{eqnarray}
  {\bf B}_{\rm p} \times ({\bf v}_{\rm p} \times {\bf B}_{\rm p}) &=&
  {\hat \eta} \ {\bf B}_{\rm p}  \times (\nabla \times {\bf B}_{\rm p})
  \nonumber \\
  \Leftrightarrow \quad B_{\rm p}^2 \ v_\perp \ {\bf e}_{\perp} &=&
  {\hat \eta} \ {\bf B}_{\rm p}  \times \left(\nabla B_{\rm p}
    \times {\bf   e}_{\parallel} + B_{\rm p} \nabla \times {\bf e}_{\parallel}
  \right) \nonumber \\
  \Rightarrow \quad \quad v_{\perp} \ {\bf e}_{\perp} &=&
  {\hat \eta} \ \left( {\bf e}_{\perp} \cdot \nabla (\ln |B_{\rm p}|)
    - (\nabla \times {\bf e}_{\parallel}) \cdot {\bf e}_{\rm t} \right)
  {\bf e}_{\perp} \ .
  \label{vperp}
\end{eqnarray}

\subsection{Discussion of Inflow Properties}
\label{disc_inflow}
From the ``inflow equation'' (\ref{vperp}), the following flow properties are
evident. First, the flow magnitude is proportional to ${\hat \eta}$ and thus,
as expected, vanishes as soon as ideality is restored. Second, both magnitude
and direction of ${\bf B_{\rm p}}$ play no role for ${\bf v}_\perp$ or, in
other words, a substitution
${\bf B_{\rm p}} \rightarrow \pm \alpha \ {\bf B_{\rm p}}$ leaves
${\bf v}_\perp$ unchanged for any constant $\alpha$. (Note that a substitution
${\bf B_{\rm p}} \rightarrow - {\bf B_{\rm p}}$ changes the direction of both
${\bf e}_\parallel$ and ${\bf e}_{\perp}$, thereby preserving the direction of
${\bf v}_\perp$.)\\
Moreover, since for a flux tube $B_{\rm p}$ generically decreases outwards,
there will generally be an {\em inflow} of matter into the tube throughout the
entire region where $\eta \neq 0$ due to the first term on the rhs of Equation
(\ref{vperp}). The contribution of the second term will be negligible in
generic cases for the following reason. Both terms define a characteristic
length scale. For the first term, this is the scale ${R}$ on which the
poloidal field decreases markedly. It can be used for defining the radius of
the flux tube as well. The second term defines a typical curvature radius
${R}_c$ of the poloidal field lines. If ${R}_c$ is of the order of ${R}$, the
flux tube is strongly distorted, i.e.~the change of its cross section is of
the same size as the cross section itself. A closer analysis shows that the
field lines of ${\bf B}_{\rm p}$ have to be bent strongly inwards for the
second term to contribute to an outward directed flow and to dominate over the
first term. (For an instructive example see the Appendix.) However,
observational evidence suggests that a flux tube's cross section either stays
more or less constant (\inlinecite{Klim},\inlinecite{WaKl}) or increases
monotonously with height, as in sunspots. Noticeable amounts of inward
curvature are produced in neither of the two cases, and the second term of
(\ref{vperp}) can thus be ignored without much loss of generality. (Note that
even if such cases should occur, the notion of a tube-shaped configuration
requires that strong inward curvature of field lines at one tube part be
balanced by a suitably strong {\em outward} curvature at some other part.
Consequently, the weaker the inflow gets at one point, the stronger it gets at
some other point, as can clearly be seen in Figure 8 of the Appendix. Although
the fact that the net value of this mutual cancelling of flow depends on the
{\em global} density structure makes a precise quantitative treatment of the
most general case more difficult, it seems reasonable to assume that even
then the net inflow will be diminished only moderately by strong poloidal
curvature.)\\
In the case of straight flux tubes, the approximation of small
$\nabla \times {\bf e}_{\parallel}$ becomes exact and leads to a scaling of
the inflow velocity  
\begin{equation}
  \| {\bf v}_{\perp} \| \propto 1/{R} \ ,
\end{equation}
which means that the inflow is more violent for thinner tubes.  For instance,
comparing cylindrical flux tubes with the same ${\bf B}_{\rm p}$ profile but
different characteristic radii $R$
\begin{equation}
  {\bf B}_{\rm p} (r) = B_0 \ b_z (r/{R}) \ {\bf e}_{z}
\end{equation}
we find
\begin{equation}
  {\bf v}_{\perp} = {\hat \eta} \ \nabla (\ln |B_{\rm p}|) 
  = {\hat \eta} \ R^{-1} \ \partial_x (\ln |b_z(x)|)
\end{equation}
where the dimensionless radial coordinate $x:=r/R$ has been introduced. We may
thus conclude that the total mass inflow through a cylindrical surface of
radius $R$ occurring within the resistive layer,
\begin{eqnarray}
  \label{Mdot}
  \dot M &:= & \int (\rho \ {\bf v}_{\perp}) \cdot {\bf da}
  = \int (\rho \ v_{\perp}) \ 2 \pi R \ {\rm d}z \\ 
  & = & \ 2\pi \ \Big( \partial_x \ \ln |b_z(x)|\Big)\bigg|_{x=1}
  \int \rho(z) \ {\hat \eta(z)}  \ {\rm d}z \ ,
\end{eqnarray}
is {\em scale-independent with respect to $R$} under the assumption of a
horizontally stratified atmosphere ($\rho=\rho(z)$ and
${\hat \eta} = {\hat \eta}(z)$), i.e.\ tubes of various radii but with the
same $b_z$ profile will transport the same mass rate, regardless of their
strength. (Note that the momentum equation (\ref{impbil_T}) was not used to
derive the preceeding results, which therefore are not limited to flow fields
satisfying $v\ll v_{A}$.)

\section{Solving for the Complete Flow Field}
\label{complete}
From now on, it will be assumed for simplicity that the tube is of cylindrical
shape, i.e.\ it does not ``fan out''. This simplifying assumption is
justified by the fact that the main effects of the resistive layer, namely the
flow of plasma into the flux tube and the decoupling of toroidal velocities
above and below this layer, are both already present in this simplified
geometry. Our assumption then translates to $B_r(r,z) \equiv 0$, so that the
solenoidality condition (\ref{maxdivB}) now reads
\begin{equation}
  0 = \nabla \cdot {\bf B} = \partial_z B_z(r,z) \ .
\end{equation}
Now consider the momentum equation (\ref{impbil_T}) in its $[r,\phi,z]$
components:
\begin{equation}
  \label{imp_comp}
  \left( \begin{array}{c} 0\\0\\0\end{array}\right)=-
  \left( \begin{array}{c} \partial_r P(r,z)\\0\\
      \partial_z P(r,z)\end{array}\right) + {\bf j}\times {\bf B}+
  \rho(r,z) \left( \begin{array}{c} 0\\0\\-g\end{array}\right)
\end{equation}
From the $\phi$-component we derive $\partial_z B_{\phi}(r,z) = 0$ and hence
${\bf B}={\bf B}(r)$, such that we can integrate the pressure from the
$r$-component of (\ref{imp_comp})
\begin{equation}
  P(r,z) =  p_1(r) + p_2(z) \ ,
\end{equation}
which in turn leads to $\partial_r \rho(r,z) =0 $ due to the $z$-component of
(\ref{imp_comp}). Assuming that $P$, $\rho$ and $T$ are linked by an
{\em arbitrary} equation of state $P=P[\rho,T]$, a horizontally stratified
temperature $T(z)$ yields $\partial_r P(r,z) \equiv 0$ and hence
\begin{equation}
  \label{ff_def}
  {\bf j}\times {\bf B} \equiv {\bf 0} \ .
\end{equation}
Therefore a cylindrical flux tube has to be force-free if the temperature of
the plasma is horizontally stratified.\\
Although the requirement of strictly horizontal isotherms $T=T(z)$ is clearly
not fulfilled for very thick tubes (as readily seen by the reduced intensity
observed in sunspots), we choose to adhere to this assumption not only for the
benefits of a markedly simplified analytical treatment but also for physical
reasons. Setting $\partial_r T = 0$ would be justified provided that
sufficiently strong horizontal heat transport was present. However, under the
assumption of an ideal plasma, the (radiation-dominated) influx of heat does
not suffice to heat the interior of an embedded flux tube to the ambient
temperature level because convective energy transport is inhibited by the
tube's strong magnetic field, as was first realised by \inlinecite{Bier}. But
within the non-ideal zone, convection across magnetic surfaces is well
permitted (or even enforced, see Section \ref{sect_inflow}), such that the
radial exchange of heat will be amplified significantly. The inflow effect
thus reduces the radial temperature gradient and may possibly lead to thermal
structures with $\partial_r T \ll \partial_z T$, in which case our assumption
would be clearly justified. (This condition should be easily fulfilled for
thin tubes, while thick tubes will hardly be affected by this reasoning since
it takes too long to exchange noticeable fractions of their mass contents via
the inflow effect, see Section \ref{sect_blowup}.) Moreover, the presence of
neutral gas in this region allows for additional convective energy transport
unimpeded by magnetic fields.\\
In our case, the only non-trivial component of (\ref{ff_def}) is the
$r$-component, which reduces to
\begin{equation}
  \label{ff_dgl}
  \partial_x \left(b_z^2(x)+b_{\phi}^2(x)\right) +2 \ b_{\phi}^2(x)/x=0 \ ,
\end{equation}
where again $x \equiv r/R$, $b_{\phi}(x) := B_{\phi}(xR)/B_0$, and
$b_z(x) := B_z(xR)/B_0$ were used. This equation is solved by 
\begin{equation}
  \label{ff_sol}
  b_{\phi}^2(x) = - x \ \partial_x G(x) \quad \mbox{and} \quad
  b_z^2(x) = (1/x) \ \partial_x [x^2 G(x)]
\end{equation}
for any function $G(x)$ satisfying $\partial_x G(x) < 0$ and
$\partial_x [x^2 G(x)] > 0$ \cite{Schl}. Figure 3 shows a typical solution to
(\ref{ff_dgl}).\\
\begin{figure}
  \centerline{\includegraphics*[width=4cm]{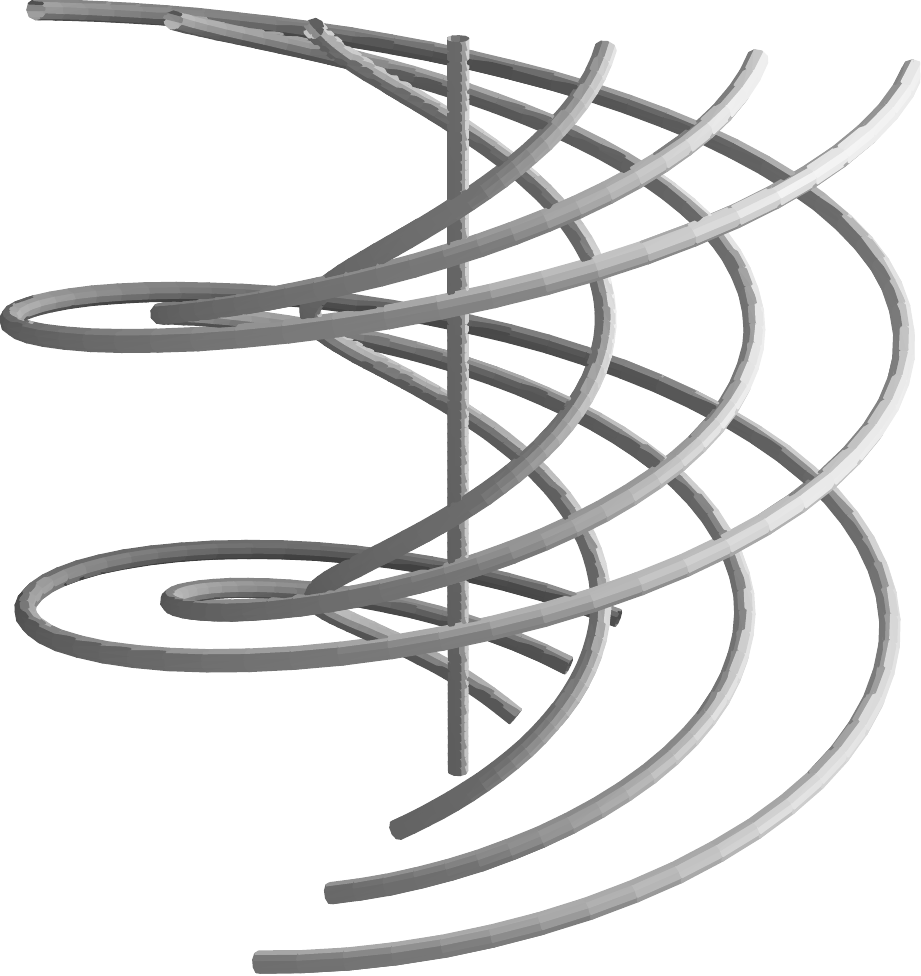}
    \caption{Selected field lines of
      $\displaystyle (b_{\phi},b_z) = \left(x/(1+x^2),1/(1+x^2)\right)$, which
      satisfies (\ref{ff_dgl}) and corresponds to $G(x)=[2(1+x^2)]^{-1}$.}
  }
\end{figure}
Given a force-free magnetic field, the solution for the flow velocity
${\bf v}$ is to be determined from Ohm's law and the equation of continuity.
This also requires to fix boundary conditions for the velocity on either the
upper or the lower boundary of the domain. Here the linearity of the two
equations with respect to ${\bf v}$ is very useful because any solution can
be seen as a superposition of a solution ${\bf v}_{\rm id}$ of the ideal Ohm's
law (\ref{idOhm}) and the continuity equation, and a particular solution
${\bf v}_{\rm res}$ of the resistive Ohm's law (\ref{ohm}) and the continuity
equation:
\begin{equation}
  {\bf v} = {\bf v}_{\rm id} + {\bf v}_{\rm res} \ .
\end{equation}
Since we are interested in the deviation from iso-rotation due to the
resistive photospheric layer, we are free to determine ${\bf v}_{\rm res}$ for
a certain choice of boundary conditions, and the solution for any other
boundary condition is then given by adding a corresponding ideal solution.
The most simple boundary condition is setting ${\bf v} = {\bf 0}$ on the upper
boundary $z=z_{\rm up}$, such that the velocity on the lower boundary exactly
equals the difference of the toroidal velocities above and below the
photosphere, i.e.~the deviation from iso-rotation. In this case we have (the
index ``res'' on the solution is suppressed in the following):
\begin{equation}
  \label{vfield}
  \begin{array}{rrrl}
    v_r(x,z) &=&   - R^{-1}&        \beta_1(x) \ \eta(z) \\
    v_{\phi}(x,z) &=& \pm R^{-2}& \big[ \ \beta_2(x) \ I_2(z)
    + \beta_3(x) \ I_1(z) \ \big] \\
    v_z(x,z) &=&   + R^{-2}& \beta_4(x) \ I_2(z)
  \end{array}
\end{equation}
where the $\beta_{1...4}(x)$ are given by
\begin{eqnarray}
  \beta_{1}(x) &:=& \frac{b_z^{\prime}(x)}{b_z(x)} \\
  \beta_{2}(x) &:=&  \frac{b_{\phi}(x)}{b_z(x)}
  \left[ \frac{b_z^{\prime \prime}(x)}{b_z(x)}
    + \frac{1}{x} \frac{b_z^{\prime}(x)}{b_z(x)}
    - \left( \frac{b_z^{\prime}(x)}{b_z(x)} \right)^2 \right] \\
  \beta_{3}(x) &:=& \frac{b_{\phi}(x)}{b_z(x)}
  \left[ \frac{b_z^{\prime \prime}(x)}{b_z(x)}
    -\left(\frac{b_z^{\prime}(x)}{b_z(x)}\right)^2+\frac{1}{x^2} \right] + \\
  &+& \frac{b_{\phi}^{\prime}(x)}{b_z(x)} \left[
    \frac{b_z^{\prime}(x)}{b_z(x)}-\frac{b_{\phi}^{\prime \prime}(x)}{b_{\phi}^{\prime}(x)}-\frac{1}{x} \right] \nonumber \\
  \beta_{4}(x) &:=& \frac{b_z^{\prime \prime}(x)}{b_z(x)} +\frac{1}{x}
  \frac{b_z^{\prime}(x)}{b_z(x)}
  -\left( \frac{b_z^{\prime}(x)}{b_z(x)} \right)^2  
\end{eqnarray}
and the $I_{1,2}(z)$ are defined as
\begin{equation}
  \label{defI12}
  I_1(z) := \int_z^{z_{\rm up}} \eta(\zeta) \ {\rm d}\zeta \quad \mbox{and} \quad
  I_2(z) := \int_z^{z_{\rm up}} \eta(\zeta) \ \frac{\rho(\zeta)}{\rho(z)}
  \ {\rm d}\zeta \ . 
\end{equation}
The sign of $v_{\phi}(x,z)$ in (\ref{vfield}) is opposite to the sign of
$b_{\phi}(x)$, which is not fixed by (\ref{ff_dgl}) and may be chosen
arbitrarily. (Figure 3 has sgn $b_{\phi} = +1$.)\\
To proceed further, one could either prescribe a vortex at $z=0$ and use
(\ref{ff_dgl}) and (\ref{vfield}) to compute the magnetic field components, or
insert into (\ref{vfield}) a typical solution of (\ref{ff_dgl}). The first
alternative would use a (rather long and messy) first order ODE, while in the
latter case the flow field could be read off directly from (\ref{vfield}).
Therefore, this avenue is chosen here.

\section{Realistic Input Parameters}
For a quantitative evaluation, prescription of density and resistivity
profiles $\rho(z)$ and $\eta(z)$ is required. We use the data provided by the
solar atmosphere model ``C'' of \inlinecite{Vern} (hereafter VAL, see
Figure 4) along with the conductivity calculations by \inlinecite{KuKa} based
on the VAL model. Since this model neglects magnetic forces, we continue to
assume isotropic resistivity for both simplicity and consistency, that is, we
take $\eta \equiv (\sigma_{\parallel})^{-1}$ from Table III in \cite{KuKa}
(see Figure 5). The function
\begin{equation}
  \label{etafunc}
  \eta(z) =
  \eta_0 \ \left[1+\left(\frac{z-z_{\rm m}}{L_{\eta}}\right)^4\right]^{-1}
\end{equation}
with [$\eta_0 = 0.058 \ \Omega$ m, $z_{\rm m} = 360$ km, $L_{\eta} = 330$ km]
also depicted there will be used to model the photosphere's actual
resistivity; its density is approximated by
\begin{equation} 
  \rho(z) = \rho_0 \ \exp (-z/L_{\rho})
\end{equation} 
with $\rho_0 = 3 \cdot 10^{-4} \ {\rm kg \ m^{-3}}$ and
$L_{\rho} = 120 \ {\rm km}$, which is in sufficient agreement with the VAL
model data for $\rho(z)$.
\begin{figure}[H]
  \centerline{ \includegraphics[width=10cm]{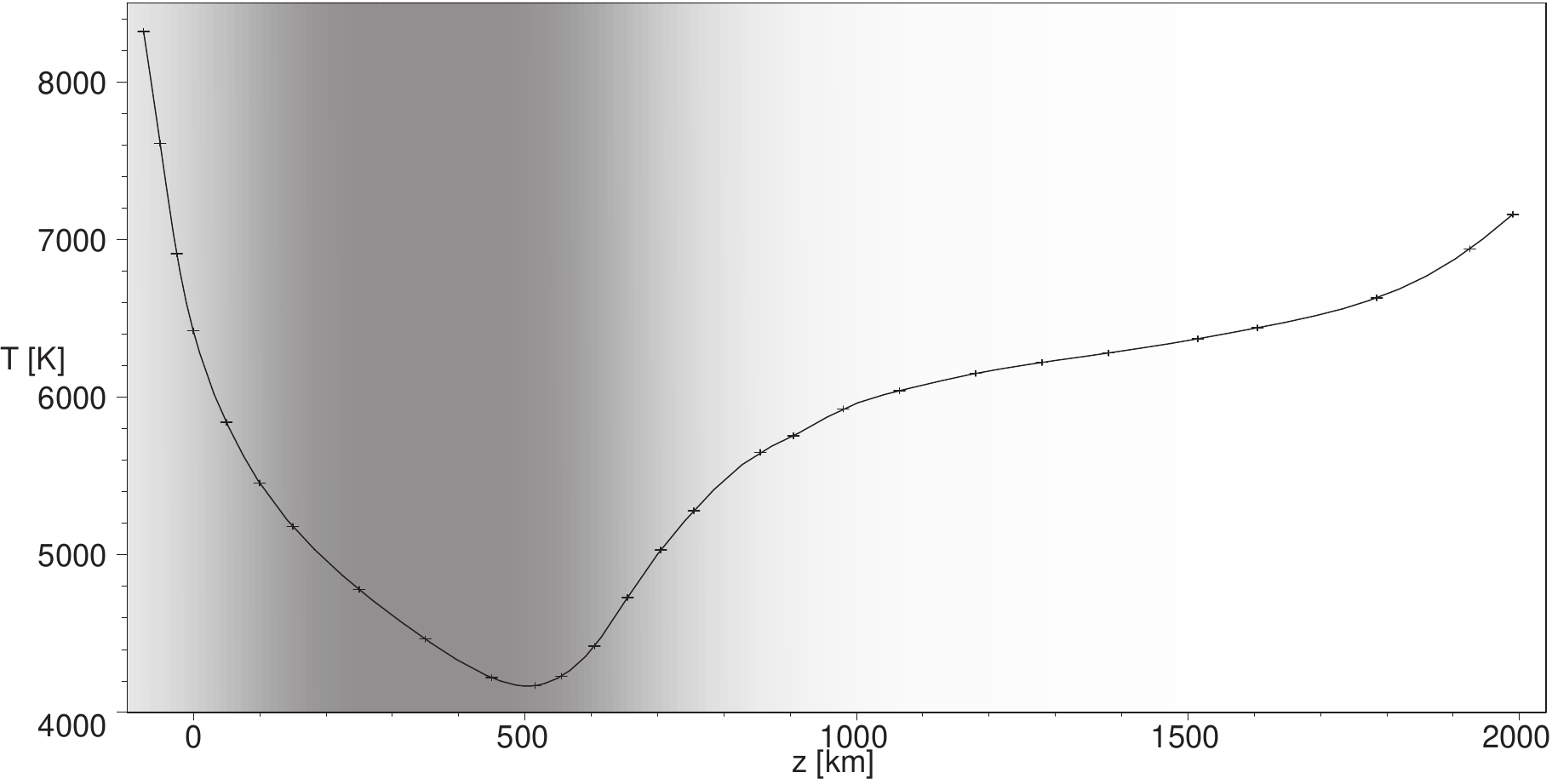}
    \caption{Temperature variation with height according to the VAL model. The
      shaded area marks the resistive region, in which the ionisation ratio
      drops below $10^{-5}$. The minimum of $T(z)$ at $z \approx$ 500 km is
      clearly discernible.}
  }
\end{figure}
\begin{figure}[H]
  \centerline{
    \includegraphics[width=10cm]{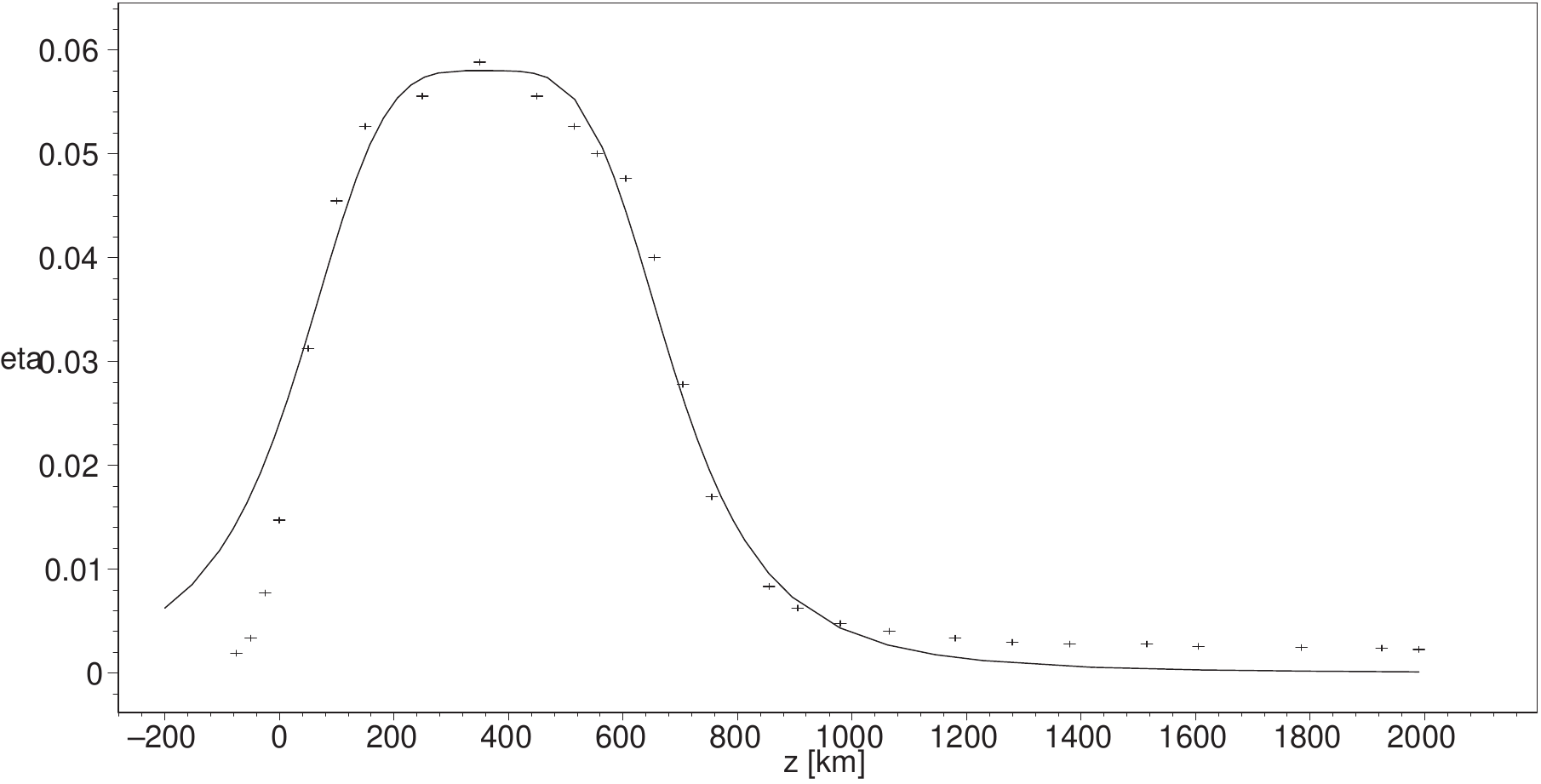}
    \caption{Computed resistivity profile (crosses) vs. analytic model
      function. The relatively poor agreement between relation (\ref{etafunc})
      and the data at $z<0$ is of minor importance since these layers are not
      explicitly considered here.}
  }
\end{figure}
\noindent
In the ensuing quantifications, we will specialise to the B field of Figure~3
as a ``flux tube prototype''. This seems justified since tentative
computations using other fields have yielded only very small deviations.
Additionally, $z_{\rm up} \rightarrow \infty$ is used since above the non-ideal
region the contribution to (\ref{defI12}) becomes negligible.
\newpage

\section{Quantitative Flow Evaluation}
\label{quantify}
\subsection{The Scaling Law}
According to (\ref{vfield}), there must be a tube radius $R_{\rm tr.}$ such that
\begin{equation}
  v \equiv \|{\bf v}\| \propto \left\{
    \begin{array}{ccc}
      R^{-1} & : & R \gg R_{\rm tr.} \\
      R^{-2} & : & R \ll R_{\rm tr.}
    \end{array}\right.
\end{equation}
Inserting our parameters found in the preceeding section, we find
$R_{\rm tr.} \approx$ 5000 km. Since at this radius $v$ will have dropped
below 1 m s$^{-1}$, we may safely regard
\begin{equation}
  \|{\bf v}\| \propto R^{-2}
\end{equation}
as the relevant scaling law for small scale flux tubes.

\subsection{The Blowup Timescale}
\label{sect_blowup}
The knowledge of absolute photospheric density and resistivity allows us to
quantify the total mass inflow (\ref{Mdot}) associated with a cylindrical tube
as $\dot M \approx 1.9 \cdot 10^8$ kg s$^{-1}$, which, when compared to the
total mass
\begin{equation}
  M_{\rm tot} := \int_{r<R} \rho \ {\rm d}V
  = \pi R^2\int_0^{z_{\rm up}} \rho(z) \ {\rm d}z
  \approx 4.9 \cdot 10^{13} \ {\rm kg} \left(\frac{R}{100 \ {\rm km}} \right)^2
\end{equation}
of the plasma contained inside the tube, defines a typical timescale
\begin{equation}
  \tau_{\rm blowup} := \frac{M_{\rm tot}}{\dot M} = \frac{R^2}{2} \
  \frac{\displaystyle\int_0^{z_{\rm up}} \rho(z) \ {\rm d}z}{\displaystyle\int_0^{z_{\rm up}} \rho(z) \ {\hat \eta}(z) \ {\rm d}z} \approx 70 \ {\rm h} \left(\frac{R}{100 \ {\rm km}}\right)^2
\end{equation} 
at which the tube exchanges a noticeable fraction of its contents.

\subsection{Condition for sub-Alfv\'enic Flows}
\label{subA-flows}
Since the flow magnitude scales $\propto R^{-2}$, the requirement
$v \ll v_{\rm A}$ is actually a limitation on the radius of the flux tube.
According to Figure 6, this may be quantified as $R \gsim 10$ km, which is
well below the resolvable scale achieved by present (and near-future) solar
observations. Discarding the inertia term in (\ref{impbil_T}) was indeed
justified.

\begin{figure}[H]
  \includegraphics[width=10cm]{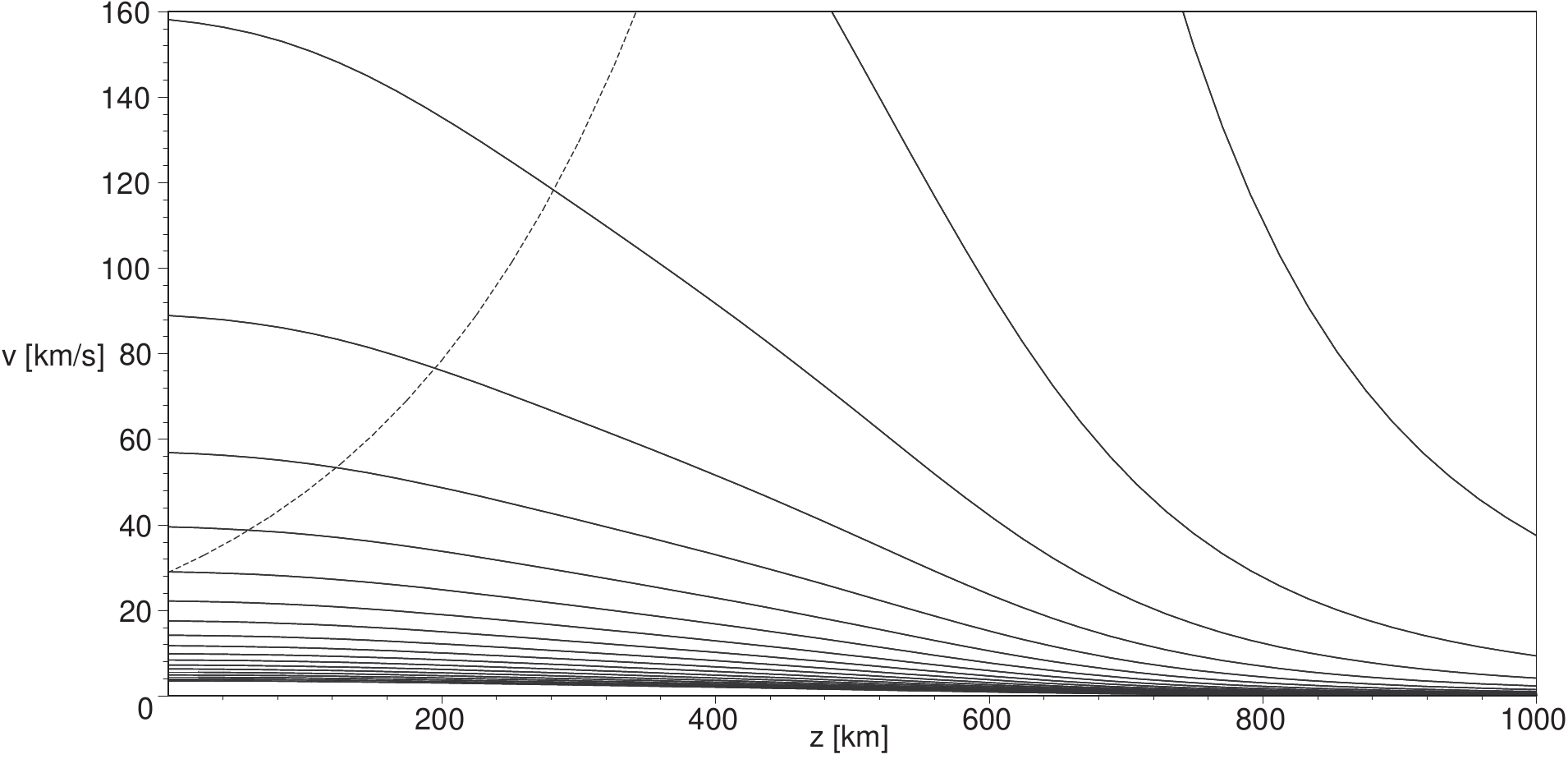}
  \caption{Maximum flow magnitude for tube radii
    $R \in \{1, 2, ...20 \ {\rm km}\}$ (solid) vs. Alfv\'en speed (dashed).
    $v_{\rm A}$ increases $\propto [\rho(z)]^{-1/2}$, i.e.~exponentially.}
\end{figure}
\begin{figure}[h]
  \includegraphics[width=10cm]{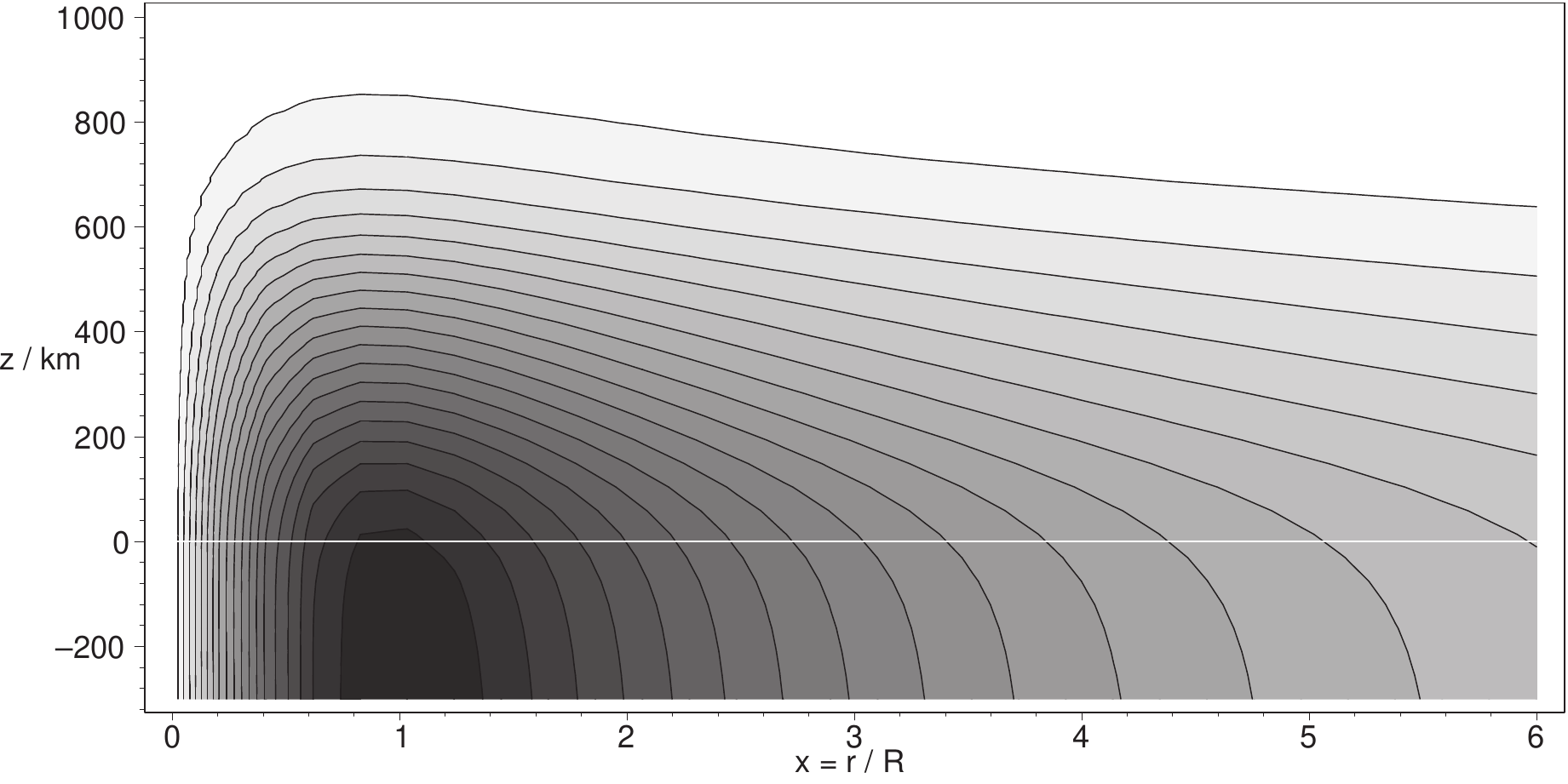}
  \caption{Poloidal contour plot of $v_{\phi}(x,z)$.}
\end{figure}
\subsection{Field Line Slippage}
The toroidal flow depicted in Figure 7 shows a striking deviation from the
flow expected in the ideal case $\eta =0$ (in which Ferrano's theorem of
{\em iso-rotation} forces all field lines to rotate with constant angular
velocity $\Omega(x,z) :=v_{\phi}(x,z)/(xR)$, such that
$\partial_z \Omega \equiv 0$). Note also that far below the photosphere
($z<0$), the lines of constant $v_{\phi}$ tend to become vertical, implying
that iso-rotation is recovered as $\eta$ again tends to zero.\\
Since we have $I_2(0)/I_1(0) \approx 0.1 \ll 1$, the profile of the footpoint
vortex (i.e.~the cut along $z=0$) can be approximated by
\begin{equation}
  \frac{v_{\phi}(x,0)}{130 \ {\rm m \ s^{-1}}} \approx
  \left(\frac{R}{100 \ {\rm km}}\right)^{-2} \frac{x}{1+x^2} \ .
\end{equation}
With $\Omega(x,z_{\rm up})=0$, the total difference in angular velocity below
and above the non-ideal layer is given by
\begin{eqnarray}
  \Delta \Omega(x) &:=& \frac{|v_{\phi}(x,0)|}{xR} \approx \frac{4}{R^3} \
  \frac{1}{1+x^2} \ \int_0^{z_{\rm up}} {\hat \eta}(\zeta) \ {\rm d}\zeta\\
  &<& \Delta \Omega(0) \approx 3.1 \cdot 10^{-4} \ {\rm s}^{-1} \
  [R/(100 \ \rm km)]^{-3} \ .
\end{eqnarray}
For observable tube sizes ($R \gsim 100 \ {\rm  km}$) this would require a
velocity resolution close
to $(\Delta v)_{\rm reso} := R \ \Delta \Omega \approx 30 \ {\rm m \ s^{-1}}$.
Although this limit is not quite reached by current imaging techniques, the
further improvement in image resolution may soon render observational
verification feasible. 

\section{Implications for the Tube's Global Evolution}
Since, according to our results, plasma has to flow into the tube from both
ends and cannot leave the tube outside the non-ideal zone, the question arises
as to where the inflowing matter goes. The possibilities are a) a steady
increase of the tube's volume (tube gets ``inflated''), b) tube is static and
downflow into the convection zone occurs or c) the inflowing plasma recombines
within the flux tube and leaves the tube in the form of neutral gas. (Of
course, in reality various combinations of a) to c) are conceivable.) In the
present model, the direction of vertical flow inside the tube is determined by
the boundary condition at $z=z_{\rm up}$ (or any other height), such that up-
or downflows of arbitrary magnitude can be achieved by choosing a
correspondingly large (possibly negative-valued) profile for
$x \mapsto v_z(x,z_{\rm up})$. However, we have no reason to favour any
specific profile, and thus our present, rather simple model cannot provide a
definitive answer here. (Note that in Section \ref{complete},
${\bf v}|_{z_{\rm up}}={\bf 0}$ (implying downflow at $z=0$) was merely chosen
to simplify the calculation of field line slippage. It was not supposed to
indicate a preference for downflows in any way.) To shed light on this
important issue, the aforementioned possibilities a) to c) suggest two avenues
for an extension of our model. First, if plasma was flowing up the tube,
thereby forcing it to expand in length and/or cross section, the tube's field
lines would be stretched, and their tension increased. Eventually, the growing
contribution from the ${\bf j \times B}$ force might become strong enough to
balance the gas pressure, causing the upflow to cease. To see whether such a
final equilibrium state exists, and if so, what the tube parameters in such a
state are, one would need to abandon cylindrical symmetry and model the full
arch-shaped tube such that all effects of field line curvature could properly
be accounted for. Unfortunately, the corresponding set of equations could turn
out to be very difficult to solve analytically, and thus the feasibility of
this approach is unclear at the moment.\\
Second, evaluating the significance of possibility c) would obviously require
the introduction of radial gradients of ionisation and temperature. In such a
model, one-dimensional reference atmospheres such as VAL can no longer be used
to prescribe atmospheric parameters (except at large distances from the tube
axis), and  self-consistent modelling of density and temperature becomes
mandatory. Again, it seems doubtful whether analytic solutions can be obtained
at a reasonable expenditure.\\
Still, in both cases a recourse to numerical investigations of the described
settings remains a vital option and may help to clarify the role of the inflow
effect with respect to the tube's global temporal evolution. (A discussion of
observational evidence for downflow is given by \inlinecite{Frut}, but whether
these observations can be applied to the photospheric region remains unclear
since they refer to velocities measured at coronal or transition region
temperatures. Generally speaking, the very existence of pronounced vertical
flows inside photospheric flux tubes still seems to be a controversial issue
among the observing community.) 

\section{Summary}
Our analytic investigation of stationary MHD equilibria of magnetic flux tubes
has shown that Ohm's law enforces an inflow of fluid towards loci of higher
field strength, which depends neither on the tube's cross section, nor on the
strength and direction of its {\bf B} field. Being proportional to $\eta$,
this inflow occurs wherever the tube penetrates the cool photospheric layer,
in particular at the tube's footpoints.\\
It was shown that a static flux tube of cylindrical shape has to be force-free
if the ambient plasma temperature is horizontally stratified, a result which
holds for arbitrary values of plasma beta. The introduction of a resistive
layer allows for stationary MHD solutions with finite field twist and a
difference in the rotational velocity above and below this resistive layer.
This constitutes a marked deviation from the iso-rotational behaviour known
from ideal MHD and limits the winding-up of the flux tube's field lines if
incompatible rotational velocities are imposed on the tube's footpoints.
Although according to the scaling law for the plasma flows these effects
either too small or too slow to be detected by present solar observations
(i.e.\ the effect either requires too small structures or produces velocities
below the detection threshold), a future improvement of observational
resolution may soon show whether the described effects can be distinguished
from the convective motions of the ambient plasma.

\begin{appendix}
\label{spoiler}
To clarify some aspects concerning the direction of radial flow, consider
the magnetic field
\begin{equation}
  \label{spoilerfeld}
  {\bf B}_{\rm sp} := \frac{B_0 \ \cosh z}{1+(r \ \cosh z)^2}
  \bigg(-r \ \sinh z,\ 0,\ \cosh z \bigg)
\end{equation}
where the $r$ and $z$ coordinates are now dimensionless for simplicity of the
argument. The tube radii have a $(\cosh \ z)^{-1}$ profile and the field
decays as $\| {\bf B}_{\rm sp} \|_{z=0} \propto 1/(1+r^2)$. Figure 8 shows a
vector plot of the corresponding perpendicular flow component
${\bf v}_{\perp}$. Beyond the dotted line, the field curvature gets so strong
that the flow direction is indeed reversed, leading to an {\em outflow} of
plasma. Although the existence of photospheric fields like (\ref{spoilerfeld})
cannot be ruled out completely, the rather low astrophysical significance of
such configurations (as discussed in Section \ref{disc_inflow}) is further
diminished by the fact that ${\bf B}_{\rm sp}$ has
$\nabla \times ({\bf j \times B})_{\rm sp} \neq {\bf 0}$ and therefore does
not describe a flux tube in the hydrodynamical sense.

\begin{figure}[h]
  \centerline{
    \includegraphics[width=10cm]{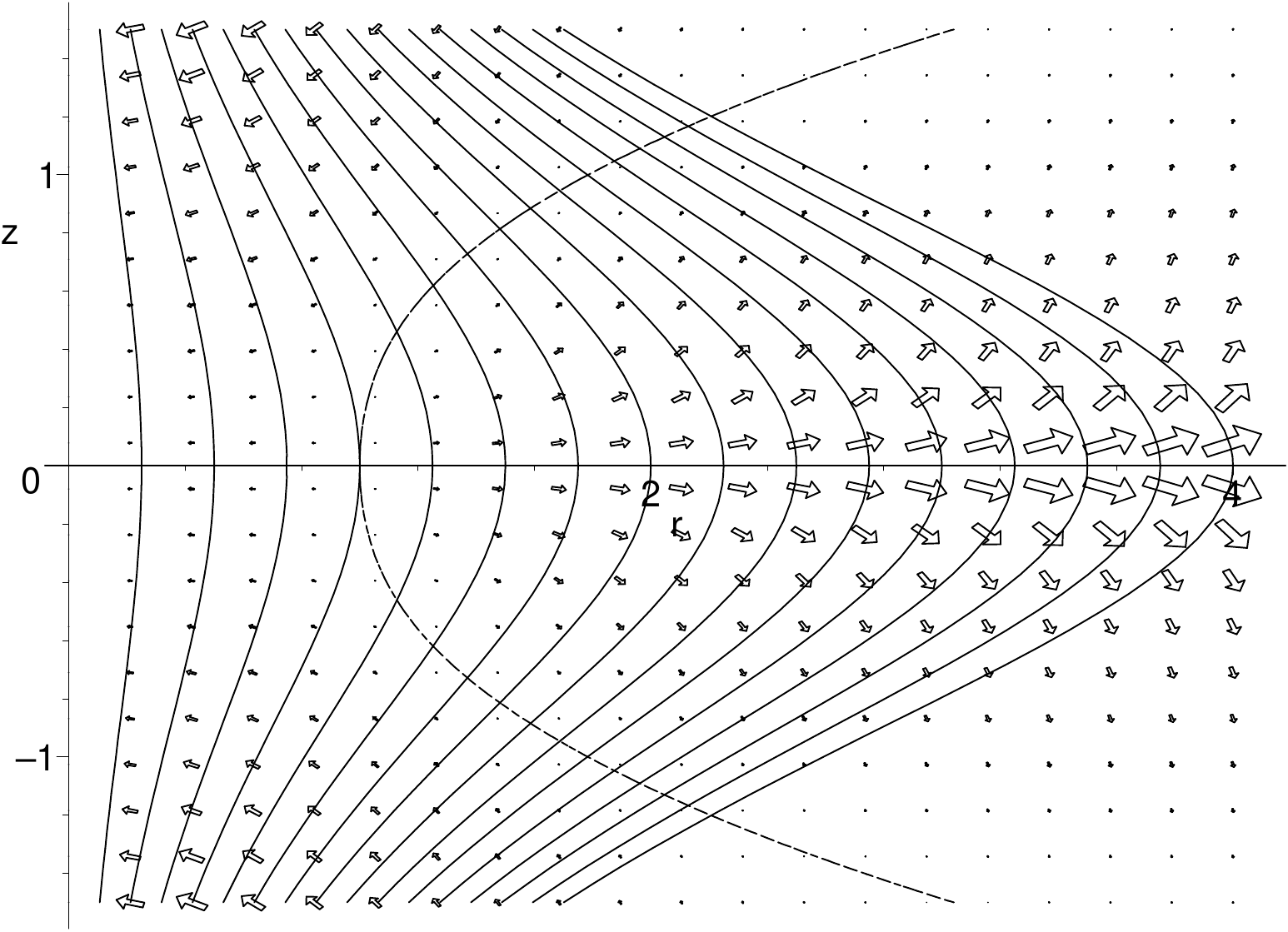}
    \caption{Vector field plot of ${\bf v}_{\perp}$ and selected field lines
      of {\bf B} (solid) in the poloidal plane. The flow direction reverses
      at the dashed line. Note how high curvature increases both inflow (due
      to the ``inflow effect'') and outflow (due to the requirement that the
      tube be void of field line reversals).}
  }
\end{figure}
\end{appendix}

\begin{acknowledgements}
  Financial support by the {\it Volkswagen Foundation} is gratefully
  acknowledged. We also thank Dr Slava Titov and the referee Dr Thomas
  Neukirch for their useful comments and Dr Vahe Petrosian for providing
  references regarding the observation of coronal loops. 
\end{acknowledgements}

\end{article}

\begin{thebibliography}
  
\bibitem[\protect\citeauthoryear{Biermann}{1941}]{Bier}
Biermann, L.: 1941, Vierteljahresschr. Astr. Ges., {\bf 76}, 194

\bibitem[\protect\citeauthoryear{Frutiger and Solanki}{1998}]{Frut}
Frutiger, C. and Solanki, S.K.: 1998, {\it Astron. Astrophys.}, {\bf 336}, 65.

\bibitem[\protect\citeauthoryear{Klimchuk}{2000}]{Klim}
Klimchuk, J.A.: 2000, {\it Solar Phys.}, {\bf 193}, 53.

\bibitem[\protect\citeauthoryear{Kub\'at and Karlick\'y}{1985}]{KuKa}
Kub\`at, J. and Karlick\`y, M.: 1986, {\it Astr. Instit. of Czechoslovakia, Bulletin}, {\bf 37}, 155.

\bibitem[\protect\citeauthoryear{Moffat}{1978}]{Moff}
Moffat, H. K.: 1978, {\it Magnetic Field Generation in Electrically Conducting Fluids}, Cambridge University Press, Cambridge, p. 65. 

\bibitem[\protect\citeauthoryear{Schl\"uter}{1957}]{Schl}
Schl\"uter, A.: 1957, {\it Z. Naturf.}, {\bf 12a}, 855.

\bibitem[\protect\citeauthoryear{Vernazza, Avrett, and Loeser}{1981}]{Vern}
Vernazza, J., Avrett, E., and Loeser, R.: 1981, {\it Astrophys. J. Suppl.} {\bf 45}, 635.

\bibitem[\protect\citeauthoryear{Watko and Klimchuk}{2000}]{WaKl}
Watko, J.A. and Klimchuk, J.A.: 2000, {\it Solar Phys.} {\bf 193}, 77.

\end{thebibliography}
\end{document}